\documentclass{INTERSPEECH2023}

\interspeechcameraready

\title{Optimizing Nepali PDF Extraction: A Comparative Study of Parser and OCR Technologies}
\name{Prabin Paudel$^*$, Supriya Khadka, Ranju G.C., Rahul Shah, Basanta Joshi}

\address{
  Department of Electronics and Computer Engineering\\Institute of Engineering, Pulchowk Campus, Nepal}
\email{075bct060.prabin@pcampus.edu.np}

\begin{document}

\maketitle
 
\begin{abstract}
This research compares PDF parsing and Optical Character Recognition (OCR) methods for extracting Nepali content from PDFs. PDF parsing offers fast and accurate extraction but faces challenges with non-Unicode Nepali fonts. OCR, specifically PyTesseract, overcomes these challenges, providing versatility for both digital and scanned PDFs. The study reveals that while PDF parsers are faster, their accuracy fluctuates based on PDF types. In contrast, OCRs, with a focus on PyTesseract, demonstrate consistent accuracy at the expense of slightly longer extraction times. Considering the project's emphasis on Nepali PDFs, PyTesseract emerges as the most suitable library, balancing extraction speed and accuracy.
\end{abstract}

\noindent\textbf{Index Terms}: Parsing, Optical Character Recognition, Unicode, PyTesseract

\section{Introduction}
\makeatletter{\renewcommand*{\@makefnmark}{}
\footnotetext{*Corresponding Author}\makeatother}

\blfootnote{This research is part of a larger project focused on Nepali Text-to-Speech Synthesis\cite{khadka2023nepali} with the objective of generating audiobooks from Nepali PDFs.\cite{paudel2023shruti}}

Text extraction, the process of retrieving content from PDF documents and converting it into a desired format, plays a pivotal role in information retrieval and analysis. This research delves into a comparitive study of the two primary methods of text extraction: PDF Parsing and Optical Character Recognition (OCR), particularly for the Nepali language. While previous works have addressed the benchmarking of PDF information extraction, such as the study by Bast et al. \cite{bast2017benchmark}, these efforts primarily pertain to languages with ample resources and may not adequately address the nuances of low-resource languages like Nepali. Additionally, existing benchmarking studies for extraction tools are often confined to academic papers and primarily focus on the English language \cite{meuschke2023benchmark} \cite{boschen2018survey}. This research seeks to bridge this gap by providing a comprehensive comparison tailored to the unique challenges of Nepali PDF text extraction.

\subsection{PDF Parsing}
PDF Parsing involves the extraction of essential information from PDF documents, and a PDF Parser serves as a tool designed for parsing the necessary content from PDFs. Typically employed for extracting substantial data from either individual or batches of PDF files, PDF Parsers have the capability to discern various elements within a PDF, including text, tables, images, metadata, and hyperlinks, allowing for the extraction of specific content. Furthermore, PDF Parsers can identify different fonts embedded in the document, contributing to their speed and accuracy in reading and processing document content. This makes PDF Parsers an optimal solution for intelligently scraping data from PDFs.

Given that the parsing process is code-driven, it enables the implementation of diverse logics to extract the desired content from any location within the document. This flexibility allows for the creation of a framework capable of extracting data from multiple sources, even when the structure of the documents varies. Notable Python libraries for PDF parsing include PyMuPDF\cite{pymupdf}, PyPDF2\cite{pypdf2} and PDFMiner\cite{pdfminer}.

\subsection{Optical Character Recognition (OCR)}
Optical Character Recognition refers to the conversion of images consisting of texts or characters within it, such as handwritten or typed, into machine-encoded text. An OCR program can extract as well as re-purpose data from a scanned document, a photo of a document or any images consisting of textual data. It extracts the content of one character or  a word at a time, joins them to form a sentence, and thus generates a textual representation of the original source. 

It generally supports different file types such as jpg, jpeg, png, and pdf as well as other formats. The output stream can be a plain text stream or a group of characters. However, an improved OCR can represent the data in a layout parallel to the original input. Some of the Python libraries for OCR include PyTesseract\cite{pytessaract} and EasyOCR\cite{easyocr}.

There are 2 basic types of core OCR algorithms, which are described below:
\begin{enumerate}
    \item \textbf{Pattern Matching}
    
    Also known as \textit{pattern recognition}, \textit{image correlation} or \textit{matrix matching}, this algorithm compares an image to a stored glyph on a pixel-by-pixel basis.\cite{christodoulakis2009evaluation} This method requires isolation of the glyph from the image and the stored glyph should be similar in font as well the scale. It is highly efficient in typewritten text but doesn't work in texts whose font it is not trained on.

    \item \textbf{Feature Extraction}
    
    This algorithm decomposes glyphs into features such as lines, closed loops, line direction, and intersections.\cite{verma2012survey} These features reduce the dimensionality of the representation making the detection of the character computationally efficient. The reference characters are stored in an abstract vector-like representation which are compared with the glyph’s features. Most modern OCR software as well as intelligent handwriting recognition systems use this technique.
\end{enumerate}

\section{Methodology}

This study investigates the efficacy of two primary text extraction methods—PDF Parsing and OCR —with a specific focus on handling Nepali content. The research utilized PDF parsing libraries, namely \textbf{PyMuPDF} and \textbf{PyPDF2}, to evaluate their performance on distinct PDF types. Tests involved PDFs with Nepali Unicode characters to assess extraction accuracy (PDF 1 and 2), Unicode-incompatible fonts requiring post-extraction translation (PDF 3 and 4), and image-embedded content (PDF 5). Additionally, two widely used Python OCR libraries, \textbf{EasyOCR} and \textbf{PyTesseract}, were employed to examine their ability to handle various PDF characteristics. In total, 5 different PDFs and 4 different libraries were used in our experiment. The code and data associated with this work is publicly available\footnote{\href{https://github.com/prab205/Optimizing-Nepali-PDF-Extraction}{https://github.com/prab205/Optimizing-Nepali-PDF-Extraction}}.

The conversion process for PyTesseract involved using PyMuPDF to convert PDF pages into images, subsequently feeding these images to PyTesseract for text extraction. The extracted data, represented as a string of Nepali Unicode characters, was then stored for analysis.

\section{Result and Analysis}
\subsection{PDF Parsing vs OCR: Extraction Incompatibility}
PDF Parsing is a very fast and accurate method. Parsers can extract required elements from the PDF such as text, image, and metadata. However, if the required data is within an image, it can only extract the image but not the content within the image. Moreover, PDF Parsers are not trained in Nepali fonts. If a PDF is written in Nepali Unicode, parsers can accurately extract the content but Nepali fonts such as \textit{Preeti}, \textit{Sagarmatha}, \textit{Mangal}, \textit{Kanchan} etc are not Unicode compatible. Hence, a parser extracts the English letter correspondence of the Nepali character which in turn has to be converted to Nepali content through one-to-one mapping, increasing error on the final result. 

\begin{figure}[h]
    \centering
    \includegraphics[scale=0.35]{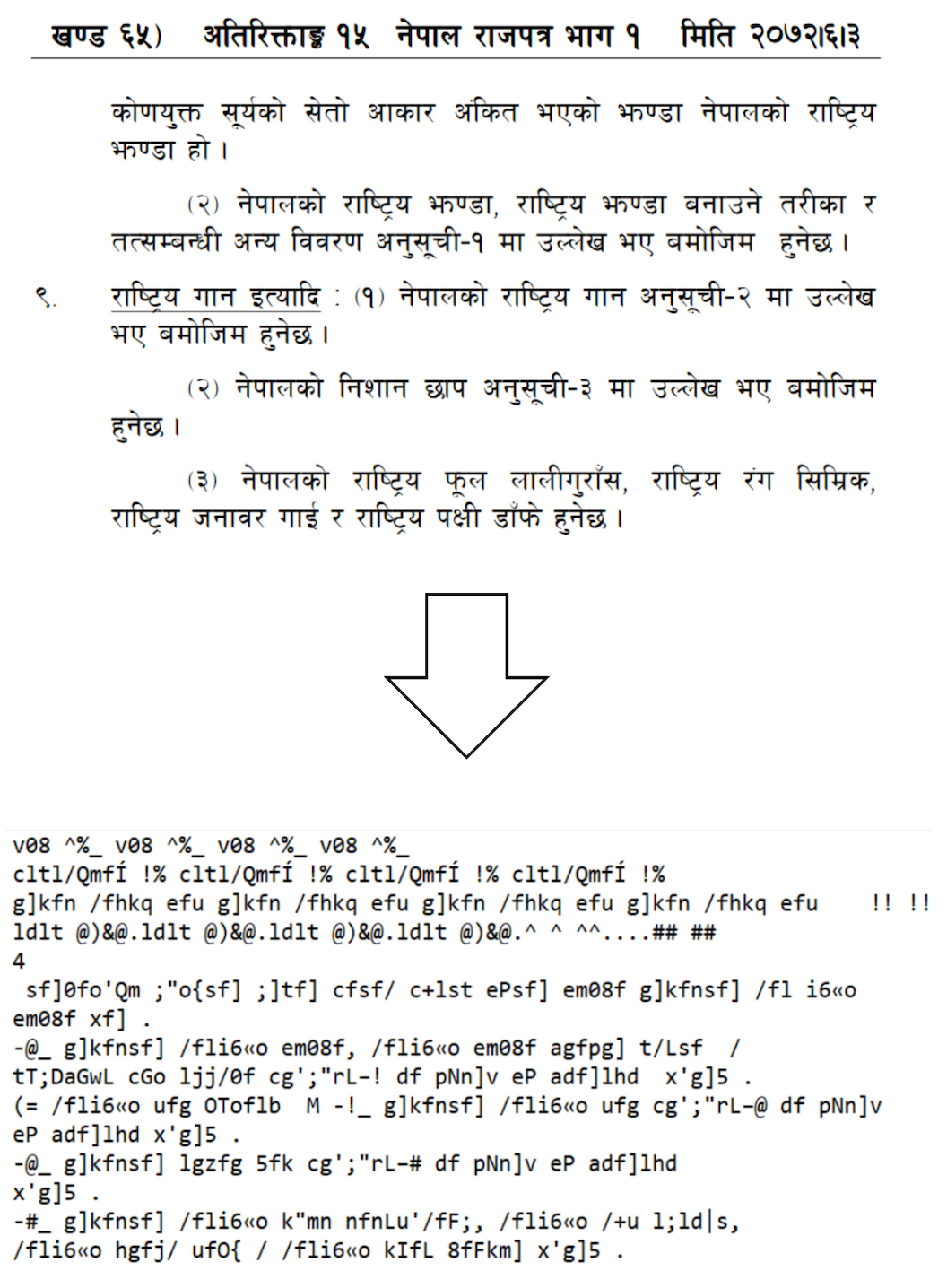}
    \caption{Extraction of Unicode Incompatible Font }
    \label{fig: Extraction of Unicode Incompatible Font}
\end{figure}
\FloatBarrier

Most of these problems are addressed by OCR which extracts the content visually. Although OCR requires an intermediate step of creating an image of the PDF, it is not limited to digitally created PDFs and can even extract the content from scanned PDFs. So, even though this method is comparatively slower than parsing, it is not dependent on the Unicode compatibility of the font as well as the type of PDF. 

EasyOCR and PyTesseract are the widely used Python OCR libraries supporting the Devanagari script. EasyOCR utilizes GPU in order to accelerate character recognition. However, without GPU, it defaults to CPU and is significantly slower than PyTesseract. 

\subsection{Time and Accuracy Comparison}

\begin{table}[htbp]
\caption{Comparison Table 1: Unicode Compatible PDFs}
\centering
\begin{tabular}{c p{0.8cm}p{1.2cm} p{0.8cm}p{1.2cm}}
\toprule
 & \multicolumn{2}{c}{\textbf{PDF 1}} & \multicolumn{2}{c}{\textbf{PDF 2}}  \\ 
 & Time & Accuracy & Time & Accuracy \\ 
 \midrule
PyMuPDF & 0.007 & 100.00 & 0.006 & 99.40\\
PyPDF2 & 0.474 & 100.00 & 0.370 & 99.26\\
\midrule
PyTesseract & 0.719 & 98.08 & 6.030 & 98.96\\
EasyOCR & 14.250 & 97.32 & 23.690 & 97.18\\ 
\bottomrule
\end{tabular}
\label{table2}
\end{table}

\begin{table}[htbp]
\caption{Comparison Table 2: Unicode Incompatible PDFs}
\centering
\begin{tabular}{c p{0.8cm}p{1.2cm} p{0.8cm}p{1.2cm}}
\toprule
 & \multicolumn{2}{c}{\textbf{PDF 3}} & \multicolumn{2}{c}{\textbf{PDF 4}}  \\ 
 & Time & Accuracy & Time & Accuracy \\ 
 \midrule
PyMuPDF & 0.008 & 96.35 & 0.009 & 86.77\\
PyPDF2 & 0.089 & 94.31 & 0.099 & 85.70\\
\midrule
PyTesseract & 1.039 & 99.81 & 1.499 & 99.81\\
EasyOCR & 17.055 & 96.51 & 19.812 & 98.97\\ 
\bottomrule
\end{tabular}
\label{table2}
\end{table}

\begin{table}[htbp]
\caption{Comparison Table 3: Image Embedded PDF}
\centering
\begin{tabular}{c p{0.8cm}p{1.2cm}}
\toprule
 & \multicolumn{2}{c}{\textbf{PDF 5}} \\ 
 & Time & Accuracy \\ 
 \midrule
PyMuPDF & - & -\\
PyPDF2 & - & -\\
\midrule
PyTesseract & 1.750 & 97.71\\
EasyOCR & 16.636 & 97.38\\ 
\bottomrule
\end{tabular}
\label{table2}
\end{table}

Different methods are compared on different types of PDFs. PDF 1 and 2 contain Nepali Unicode characters and hence PDF Parsers such as PyMuPDF and PyPDF2 have high extraction accuracy. PDF 3 and 4 are written using Unicode-incompatible Nepali fonts which have to be translated to Nepali after the extraction process, reducing the overall accuracy. The content in PDF 5 is embedded in an image. So, PDF Parsers cannot extract the content at all. Overall, the time taken by Parsers is significantly lower than the OCR but depending on the type of PDF, the output accuracy fluctuates. On the other hand, OCRs on average have greater extraction time but almost constant accuracy as shown in the graph below:

\begin{figure}[h]
    \centering
    \includegraphics[scale=0.2]{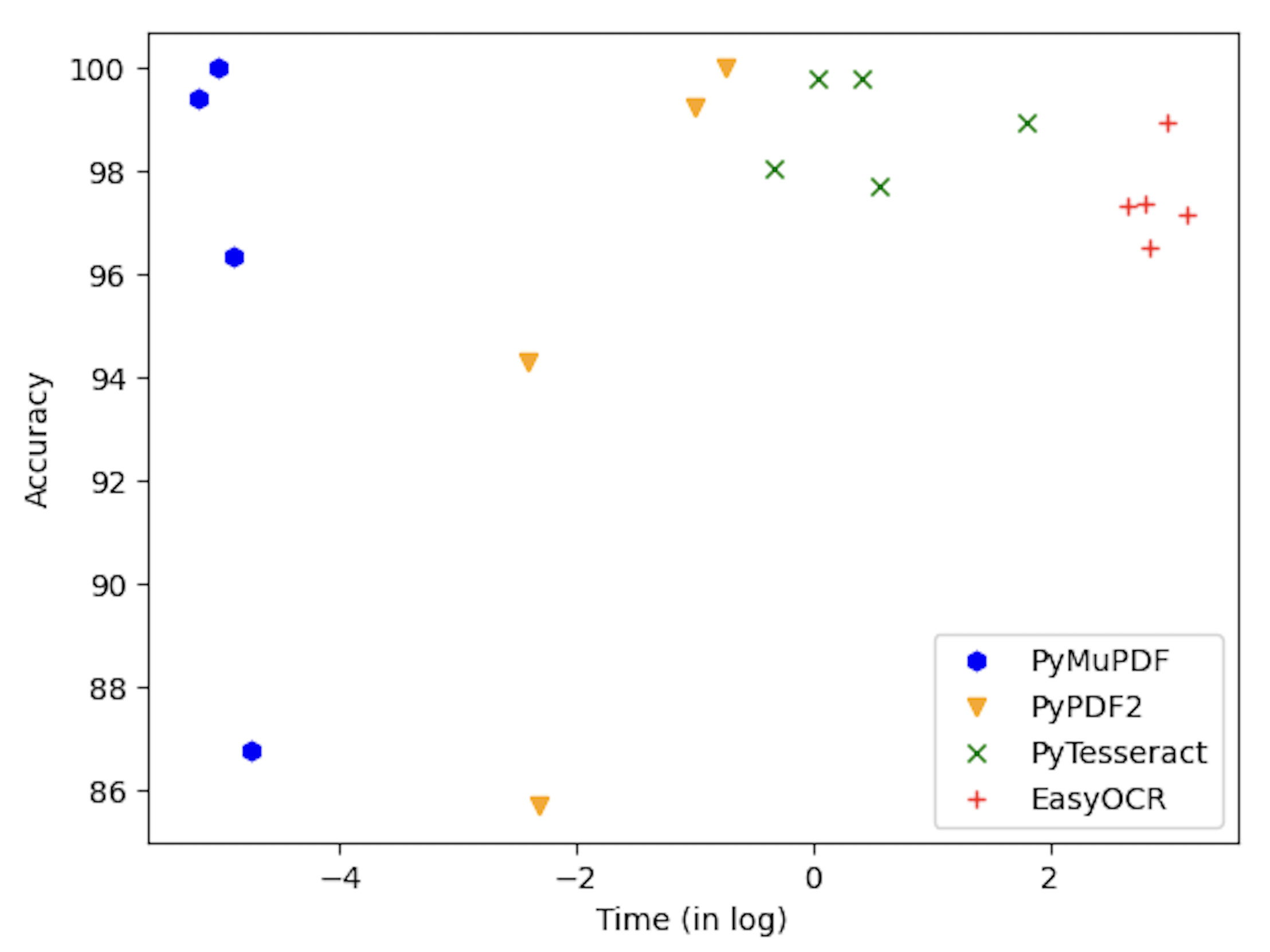}
    \caption{Comparison graph of different extraction methods }
    \label{fig: Comparison graph of different extraction methods}
\end{figure}
\FloatBarrier

Hence, weighing the time and accuracy aspect of the extraction, PyTesseract is considered to be the most suitable library for extracting the Nepali PDF. 

\section{Conclusion}
In conclusion, this research systematically compared PDF parsing and Optical Character Recognition methods for the extraction of Nepali content from PDFs. PDF parsing demonstrated commendable speed and accuracy, yet encountered challenges when dealing with non-Unicode Nepali fonts. The study highlights the effectiveness of OCR, particularly PyTesseract, in overcoming these challenges, exhibiting versatility for both digital and scanned PDFs. The trade-off between PDF parsers' faster extraction times and OCRs' consistent accuracy was analyzed, revealing PyTesseract as the most suitable library for the project's emphasis on Nepali PDFs. Notably, PyTesseract's ability to maintain accuracy across diverse PDF types outweighs the slightly longer extraction times, positioning it as the preferred choice. While EasyOCR showed promise, its comparative falter in speed makes it a close second. It is imperative to note that PDF parsers, though the fastest for digitally created Unicode PDFs, proved ineffective for extracting text embedded in images. This comprehensive evaluation provides valuable insights for selecting the optimal text extraction method, emphasizing the significance of PyTesseract in the context of Nepali PDF extraction.

\bibliographystyle{IEEEtran}
\bibliography{mybib}

\end{document}